\documentclass[conference,letterpaper]{IEEEtran}
\usepackage{enumerate}
\usepackage{multirow,array}
\usepackage{cite}
\usepackage{graphicx}
\usepackage{psfrag}
\usepackage{url}
\usepackage{amsmath}
\usepackage{amsthm}
\usepackage{array}
\usepackage{amssymb}
\usepackage{amsfonts}
\usepackage{float}

\usepackage{algorithm}
\usepackage{algpseudocode}

\usepackage{color, colortbl}
\definecolor{Gray}{gray}{0.9}

\makeatletter
\renewcommand{\boxed}[1]{\text{\fboxsep=.2em\fbox{\m@th$\displaystyle#1$}}}
\makeatother

\newcommand{\Cdb}{\mathcal{C}}
\newcommand{\p}{\pmb}
\newcommand{\Sc}{\mathcal{S}}
\newcommand{\Kt}{\bar{\mathcal{K}}}
\newcommand{\At}{\mathcal{A}}
\newcommand{\Gt}{\mathcal{G}}

\newtheorem*{conjecture*}{Conjecture}
\newtheorem{lemma}{Lemma}

\newtheorem{remark}{Remark}
\newtheorem{definition}{Definition}
\newtheorem{theorem}{Theorem}
\newtheorem{example}{Example}

\title{Coded Data Rebalancing:\\ Fundamental Limits and Constructions}

  \author{Prasad Krishnan, V. Lalitha, Lakshmi Natarajan}
  
\begin{document}
\maketitle
\begin{abstract}
Distributed databases often suffer unequal distribution of data among storage nodes, which is known as `data skew'. Data skew arises from a number of causes such as removal of existing storage nodes and addition of new empty nodes to the database. Data skew leads to performance degradations and \textcolor{black}{thus} necessitates `rebalancing' at regular intervals to reduce the amount of skew.
We define an $r$-balanced distributed database as a distributed database in which the storage across the nodes has uniform size, and each bit of the data is replicated in $r$ distinct storage nodes. We consider the problem of designing such balanced databases along with associated rebalancing schemes which maintain the $r$-balanced property under node removal and addition operations. We present a class of $r$-balanced databases (parameterized by the number of storage nodes) which have the property of structural invariance, i.e., the databases designed for different number of storage nodes have the same essential structure. For this class of $r$-balanced databases, we present rebalancing schemes which use coded transmissions between storage nodes, and characterize their communication loads under node addition and removal. We show that the communication cost incurred to rebalance our distributed database for node addition and removal is optimal, i.e., it achieves the minimum possible cost among all possible balanced distributed databases and rebalancing schemes.
\end{abstract}

\let\thefootnote\relax\footnotetext{
Dr.\ Krishnan and Dr.\ Lalitha are with the Signal Processing \& Communications Research Center, International Institute of Information Technology Hyderabad, India, email:\{prasad.krishnan,\,lalitha.v\}@iiit.ac.in.

Dr.\ Natarajan is with the Department of Electrical Engineering, Indian Institute of Technology Hyderabad, email: lakshminatarajan@iith.ac.in. 
}


\section{Introduction}
Distributed data analytics engines, such as Apache Ignite \cite{ApacheIgnite}, employ (a) a file system (such as the \textit{Hadoop File System} or HDFS \cite{HadoopFileSystem}) to distribute the data across several nodes in a cluster, and (b) a distributed computation framework (such as \textit{MapReduce}) to enable parallel processing of the distributed data. Generally, in such distributed file systems, the available data is allocated to storage nodes by splitting it into a number of chunks and storing them in the nodes with some replication factor, which also functions as a protection against node failures. For instance, in HDFS, the default replication factor is 3, which means each chunk is stored in three locations among the available nodes.

\textit{Data skew} in a cluster refers to the situation in which the data stored in the nodes is not uniformly distributed. The placement of data in the storage nodes can become skewed due to various reasons \cite{WhyDataSkew}. New nodes may arrive whenever the client running the application requires and can afford to add them, and the newly arrived nodes clearly would start off with no data in them which results in data skew. 
Existing nodes may leave due to node failures, which can be common when nodes are run on commodity hardware. 
In cloud computing frameworks, a node could also be removed from a user's database if it becomes unavailable due to excessive traffic (often due to the existence of other higher priority users that it has to serve). Such node removals may result in the reduction of \textcolor{black}{the} replication factor. Data then needs to be moved across existing nodes to reinstate the desired replication factor and this movement of data may result in data skew if not done carefully. 
The data-allocation protocol of the file system could also result in non-uniformity in storage across the nodes. Also, the client application may not uniformly add new data to the various nodes, preferring some nodes over others. The skew in the data placement in the storage nodes, which occurs because of such reasons, results in the imbalance of traffic handled by the various nodes. Nodes which possess a large quantity of data are forced to handle most of the traffic, and vice-versa. This could further result in the creation of \textit{stragglers} \cite{GoogleCloud}, which are nodes that act as a bottleneck to the completion of a distributed computing task.

In order to prevent data-skew, most distributed file systems employ a simple technique called \textit{data rebalancing} \cite{ApacheIgniteDataRebalancing,ApacheHadoopDataRebalancing,GoogleCloud,CephRebalancing}. Having detected the existence of data skew in the storage nodes based on some quantitative threshold \cite{HadoopFileSystem,HadoopDocumentation}, a data rebalancing protocol moves the data existing in the storage nodes between them so that the data skew falls below a certain threshold.  
As regular rebalancing becomes a necessity whenever there is a strong data skew, the rebalancing operations can still require transfers of huge amounts of data, especially in large clusters with 100s or 1000s of nodes, which is not uncommon in the present day. Thus, the rebalancing protocol is typically implemented in such a way that the amount of communication required to balance the nodes is kept low. However, the fundamental limits of this communication problem is not yet understood, and constructions of efficient rebalancing protocols remains open.  

In this work, we present a formal framework for the study of the rebalancing problem on distributed databases due to data skews arising from node removal and addition. We define the notion of an \textit{$r$-balanced distributed database} in which each node stores an equal fraction of the data, and the data is replicated $r$ times across the nodes. Under the instance of a node removal (also, a node addition), we give a formal definition for a rebalancing scheme which maintains the $r$ replication property across balanced databases before and after the node removal (also, node addition). 
We define the \textit{rebalancing load} as the sum of the communication load of the rebalancing schemes corresponding to both node removal and addition in a distributed database, and the \textit{optimal} rebalancing load $L^*(r)$ for a given replication factor $r$ as the minimum possible rebalancing load across all possible choices of balanced databases and rebalancing schemes.
We obtain a tight characterization of $L^*(r)$, by deriving a lower bound on $L^*(r)$ and also providing an explicit construction of balanced databases and the associated rebalancing schemes with the rebalancing load equal to the presented lower bound. 
For the case of a node failure, the rebalancing scheme we propose \textcolor{black}{makes use of} coded transmissions, hence we call our framework Coded Data Rebalancing. 
The schemes we construct enable transformation of a structured database into another equivalently structured database, i.e., keeping the initial and target balanced databases within the same class. Thus, the rebalancing schemes maintain the \textit{structural invariance} of the initial and target databases. This structural invariance enables the application of our distributed database and rebalancing schemes to any sequence of node removals and additions, and our scheme achieves the optimal communication load for each such sequence as long as the rebalancing operations are to be performed for each node removal or addition in the sequence. 

The rest of the paper is organized as follows. In Section \ref{systemmodel}, we present the system model for the rebalancing problem and the associated definitions. In Section \ref{resultssection}, we present the main result of this work (Theorem \ref{maintheoremachievabilityconverse}), which is a tight characterization of $L^*(r)$, and discuss its importance and implications. The next two sections, Section \ref{achievabilitysection} and Section \ref{sectionconverse}, are devoted to the proof of Theorem \ref{maintheoremachievabilityconverse} by showing the achievable scheme and the converse, respectively.
\section{System Model} \label{systemmodel}

\begin{figure}[t]
\centering
\includegraphics[width=0.35\textwidth]{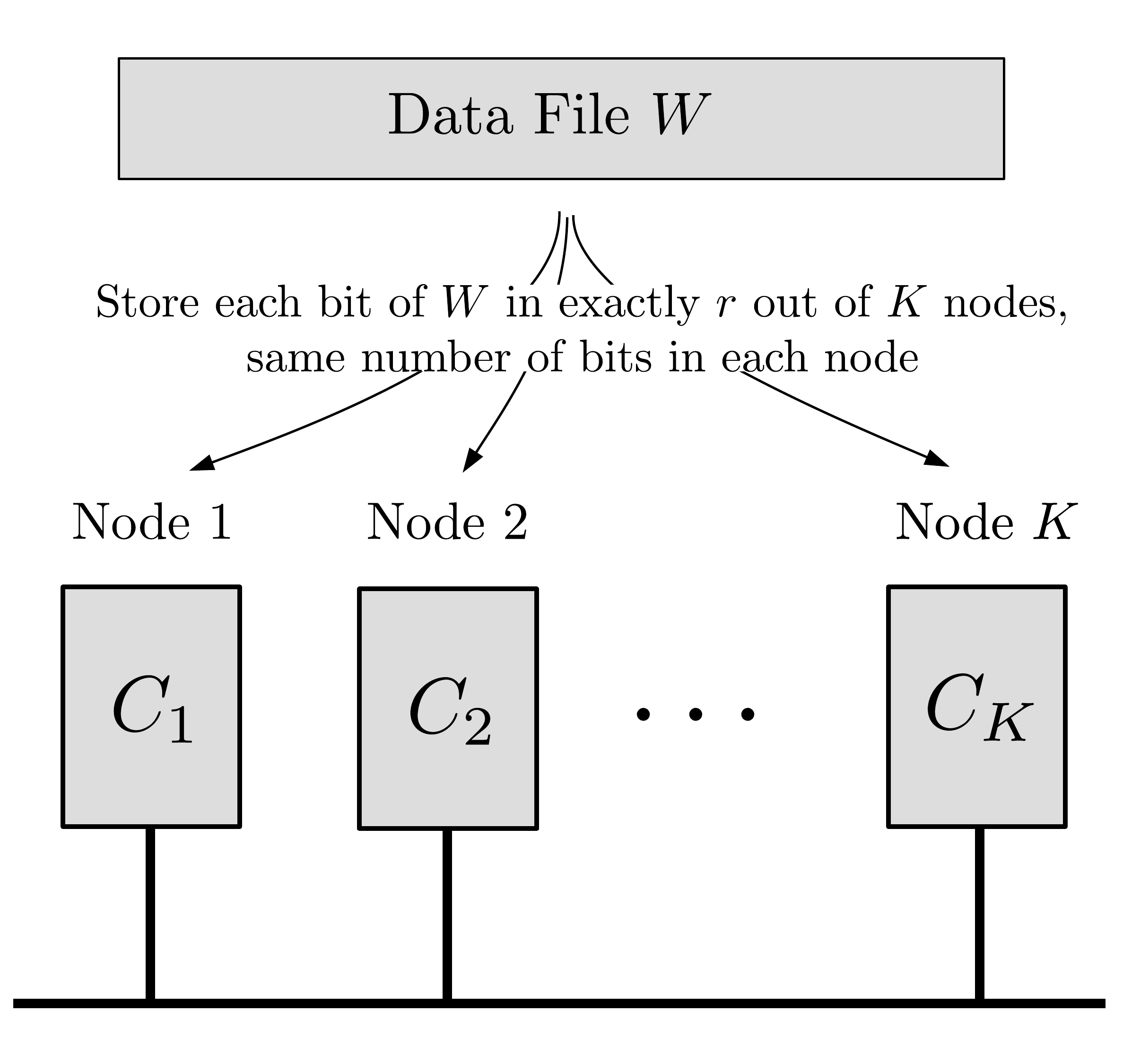}
\caption{An $r$-balanced distributed database across $K$ nodes. The storage nodes are connected by a shared communication link.}
\label{fig:distributed_database}
\end{figure}

Consider a file $W$ representing the data. We assume that the file is a set of $N$ bits with the $n^{th}$ bit denoted as $W_n\in\{0,1\}$, i.e., $W=\{W_n:n\in [N]\},$ where $[N]\triangleq \{1,\hdots,N\}.$ We consider a set of $K$ nodes, indexed as $[K]=\{1,\hdots,K\}$. The nodes $[K]$ are connected to each other via a bus link. Thus there is a noiseless broadcast channel between any node $k\in [K]$ and the other nodes $[K]\backslash k$ (the set of all elements in $[K]$ except $k$). 

A \textit{distributed database} of $W$ across nodes $[K]$ (identified by labels $[K]$) consists of a collection $\cal C$ of subsets of $W$, ${\cal C}=\{C_i\subseteq W:i\in[K]\}$ \textcolor{black}{such that $\cup_{i\in[K]}C_i=W$}, where $C_i$ denotes the set of bits of $W$ stored in node $i$. 

For a given distributed database ${\cal C}$ and a subset $B\subset[K]$, the replication factor of the $n^{th}$ bit of $W$ in $B$ is defined as
$r_n(B)=\sum_{k\in B}{\mathbb I}(W_n~\text{is stored in node}~k),$
where $\mathbb I\textcolor{black}{(\cdot)}$ denotes the indicator function.

\begin{definition}[$r$-balanced database] 
\textcolor{black}{For an integer $r\geq 1$, an $r$-balanced distributed database of file $W$ on nodes $[K]$ is a distributed database denoted by ${\cal C}(r,[K])=\{C_i \subseteq W: i \in [K]\}$},
such that 
\emph{(i)}~$r_n([K])=r, \forall n\in[N]$, and 
\emph{(ii)}~$|C_1|=|C_2|=\hdots=|C_K|$. 
We call $r$ the \underline{replication factor} of the balanced database.
\end{definition}

For any $r$-balanced distributed database, $\sum_{k\in [K]}|C_k|=rN$, and thus $|C_k|=\lambda N, \forall k\in [K]$, where $\lambda \triangleq \frac{r}{K}$ \textcolor{black}{denotes the storage fraction at any node}. 
Fig.~\ref{fig:distributed_database} illustrates the placement of the data file $W$ in an $r$-balanced distributed database.

In a given balanced distributed database \textcolor{black}{${\cal C}(r,[K])$}, the addition or removal of nodes necessitates a rebalancing operation. We now formally define the rebalancing schemes associated with node removal and addition separately, along with the communication loads associated with each. 

\subsection{Node Removal} 
Let \mbox{$k\in[K]$} be a node which is removed from the system. Let \mbox{${\cal C}_k(r,[K]\backslash k)=\{C_i(k):i\in[K]\backslash  k\}$} be a target $r$-balanced \textcolor{black}{database} that we want to achieve in the new system consisting of nodes $[K]\backslash k$. 
Let $\lambda_{rem}=\frac{\lambda K}{K-1} = \frac{r}{K-1}$. In ${\cal C}_k(r,[K]\backslash k)$, for each $j,$ we must have  $(K-1)|C_j(k)|=\sum_{i\in[K]\backslash k}|C_i(k)|=rN$. Thus, $|C_j(k)|=\lambda_{rem} N$. \textcolor{black}{Thus $\lambda_{rem}$ is the new storage fraction at a surviving node}.

\begin{figure}[t]
\centering
\includegraphics[width=0.4\textwidth]{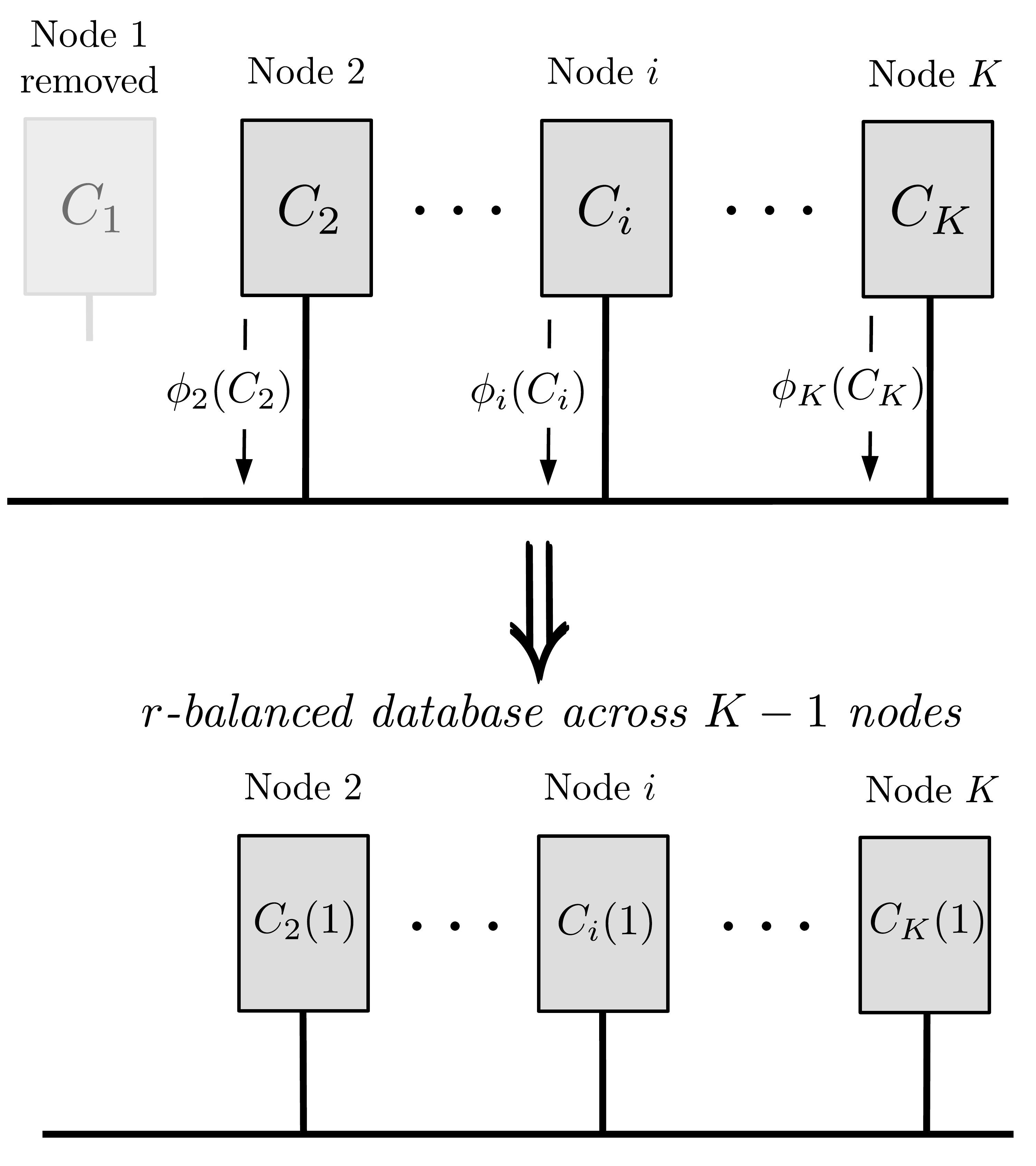}
\caption{Data rebalancing after node $1$ is removed. Each remaining node $i$ broadcasts $\phi_i(C_i)$, and uses $C_i$ and the transmissions from the other $K-2$ nodes to update its contents to $C_i(1)$.}
\label{fig:node_removal}
\end{figure}

In general, a rebalancing scheme involves each surviving storage node $i \neq k$ broadcasting a codeword $\phi_i(C_i)$ to all the other surviving nodes. At the end of these $K-1$ transmissions, each node $i \neq k$ decodes its demand \textcolor{black}{$C_i(k)$ (the storage at node $i$ in the target database ${\cal C}(r,[K]\backslash k)$)} using its current storage \textcolor{black}{$C_i$} and the received codewords using a decoding function $\psi_i$. 
Fig.~\ref{fig:node_removal} illustrates the data rebalancing operation when node $1$ is removed from the system.

\begin{definition}[Rebalancing scheme for node removal]
Let $l_i$, $i\in[K]\backslash k$, be positive integers. A \underline{rebalancing scheme} from ${\cal C}(r,[K])$ to ${\cal C}_k(r,[K]\backslash k)$, \textcolor{black}{denoted as ${\cal R}_{k,{\cal C},{\cal C}_k}\triangleq \{\phi_i,\psi_i:i\in[K]\backslash k\}$} 
, consists of a set of encoding functions $$\phi_i:\{0,1\}^{\lambda N}\rightarrow \{0,1\}^{l_i},~\text{for each}~i\in[K]\backslash k,$$ and a set of decoding functions $$\psi_{i}:\{0,1\}^{\lambda N}\times\prod_{j\in[K]\backslash\{i,k\}}\{0,1\}^{l_j}\rightarrow \{0,1\}^{\lambda_{rem}N},$$
for each $i\in[K]\backslash k$, such that 
$$
\psi_i(C_i,(X_j:j\in[K]\backslash\{i,k\}))=C_i(k),
$$ where $X_j=\phi_j(C_j)$.

The \underline{communication load} of such a rebalancing scheme is the total number of bits transmitted normalized by the number of bits of $C_k$ (the removed node's storage), given by
$$L_{rem}({\cal R}_{k,{\cal C},{\cal C}_k})\triangleq \frac{\sum\limits_{i\in[K]\backslash k}l_i}{\lambda N}.$$
\end{definition}
\subsection{Node Addition}

We now assume that a new node, indexed as $K+1$, is added to the system of nodes $[K]$. We assume that this node arrives without any data in its storage. Let ${\cal C}'(r,[K+1])=\{C_i':i\in[K+1]\}$ be a target $r$-balanced \textcolor{black}{database} to be obtained on the set of nodes $[K+1].$ Let $\lambda_{add}=\lambda\frac{K}{K+1}$. For each $j\in[K+1],$ we should have $\sum_{i\in[K+1]}|C_i'|=rN=|C_j'|(K+1)$, and thus $|C_j'|=\lambda_{add}N=\frac{rN}{K+1}$. \textcolor{black}{Thus $\lambda_{add}$ denotes the storage fraction at nodes after rebalancing.}

We assume that each $i \in [K]$ broadcasts a codeword $\phi_i'(C_i)$. The new node performs a decoding operation $\psi_{K+1}'(\phi_i'(C_i), i \in [K]) = C_{K+1}'$ using all the transmissions. Each of the nodes $i \in [K]$ decodes its demand using $C_i$ and $\phi_j'(C_j)$, $j \in [K] \setminus \{i\}$ using its own decoding function $\psi_i'(C_i,(\phi_j'(C_j), j \in [K] \setminus \{i\})) = C_i'$. See Fig.~\ref{fig:node_addition} for an illustration.
\begin{definition}[Rebalancing scheme for node addition]
Let $l_i$, $i\in[K]$, be positive integers. A rebalancing scheme from ${\cal C}(r,[K])$ to ${\cal C}'(r,[K+1])$, \textcolor{black}{denoted as ${\cal R}'_{{\cal C},{\cal C}'}\triangleq \{\phi_i',\psi_j':i\in[K],j\in[K+1]\}$}, consists of a set of encoding functions 
$$\phi_i':\{0,1\}^{\lambda N}\rightarrow \{0,1\}^{l_i},~\text{for each}~i\in[K],$$ 
and a set of decoding functions $\psi_i'$ for each $i\in[K+1]$ defined as follows.
\begin{itemize}
    \item $\psi_{i}':\{0,1\}^{\lambda N}\times\prod_{j\in[K]\backslash i}\{0,1\}^{l_j}\rightarrow \{0,1\}^{\lambda_{add}N},$ for each $i\in[K]$, such that 
$$\psi_i'(C_i,(X_j,j\in[K]\backslash\{i\}))=C_i',
\forall i\in[K]$$
\item $\psi'_{K+1}:\prod_{j \in [K]}\{0,1\}^{l_j}\rightarrow \{0,1\}^{\lambda_{add}N},$ such that $$\psi'_{K+1}(X_j,j\in[K]))=C_{K+1}',$$
\end{itemize}
where $X_j = \phi_j'(C_j)$.
\begin{figure}[t]
\centering
\includegraphics[width=0.43\textwidth]{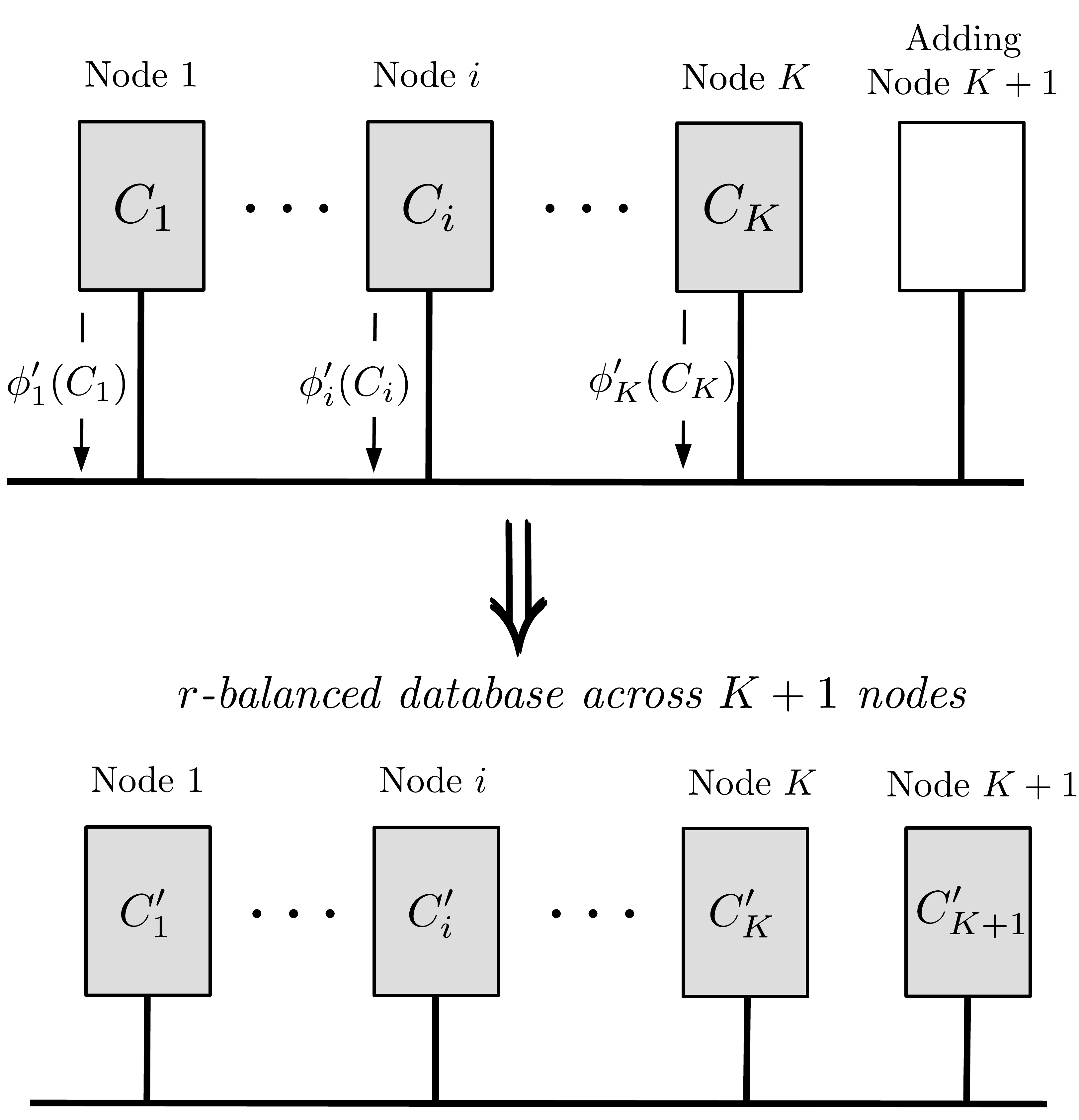}
\caption{Data rebalancing after an empty node $K+1$ is added. Each of the older nodes $i$, $1 \leq i \leq K$, broadcasts $\phi_i'(C_i)$ to the rest of the $K$ nodes. The new node uses these transmissions to construct its contents $C_{K+1}'$. The older nodes modify their contents to $C_{i}'$ using the transmissions and their current contents.}
\label{fig:node_addition}
\end{figure}

The communication load of such a rebalancing scheme is the total number of bits transmitted normalized by the number of bits $|C'_{K+1}|$ in the new node, given as
$$
L_{add}({\cal R}_{k,{\cal C},{\cal C}\backslash k})\triangleq \frac{\sum\limits_{i\in[K]}l_i}{\lambda_{add} N}.
$$
\end{definition}
\subsection{The Rebalancing Load}
\textcolor{black}{We use the sum of the loads of the node-addition and node-removal schemes as our performance metric.} 
\begin{definition}
The \underline{rebalancing load} corresponding to the rebalancing schemes, ${\cal R}_{[K]}\triangleq \{{\cal R}_{k,{\cal C},{\cal C}_k}:k\in[K]\}$ and ${\cal R}'_{{\cal C},{\cal C}'}$ as given above, is defined as
\begin{align}
\label{rebalancingloaddef}
    L({\cal R}_{[K]},{\cal R}'_{{\cal C},{\cal C}'})\triangleq \max_{k\in[K]}L_{rem}({\cal R}_{k,{\cal C},{\cal C}_k})+L_{add}({\cal R}'_{{\cal C},{\cal C}'}).
\end{align}
\end{definition}
The \textit{optimal rebalancing load} for a given replication factor $r$ is then given as the infimum of the rebalancing load (\ref{rebalancingloaddef}) over all possible choices for the balanced databases and the rebalancing schemes, i.e., 

$$L^*(r)=\inf_{{\cal C},\{{\cal C}_k:k\in[K]\},{\cal C}'}~\inf_{{\cal R}_{[K]},{\cal R}'_{{\cal C},{\cal C}'}}L({\cal R}_{[K]},{\cal R}'_{{\cal C},{\cal C}'}).$$ 
\begin{remark}
\label{remark_r_bounds_K}
Note that if the replication factor $r=1,$ then no rebalancing scheme exists for any node removal, as the fraction of the data stored in the node being removed would be irretrievably lost. Hence we always assume that $r\geq 2$. Further, if $r=K$, then maintaining this replication factor after node removal is impossible. Hence, we assume $r\leq K-1.$ 
\end{remark}
\section{\textcolor{black}{Coded Data Rebalancing}}
\label{resultssection}
The main result of this work is a tight characterization of $L^*(r)$, given by the following theorem, the proof of which is given in Section \ref{achievabilitysection} and Section \ref{sectionconverse}. 
\begin{theorem}
\label{maintheoremachievabilityconverse}
For balanced distributed databases on $K$ nodes with replication factor $r\geq 2$, the following rebalancing load $L$ is achievable
\[
L=\frac{1}{r-1}+1,
\]
if the file size $N$ is a multiple of $(r-1)\,P(K+1,K+1-r)$, where the symbol $P(K+1,K+1-r)$ denotes $(K+1)!\,/(r!)$. Further, the above load is optimal for a given replication factor $r$, i.e., $L^*(r)=\frac{1}{r-1}+1.$
\end{theorem}
\textcolor{black}{Our rebalancing schemes of Theorem \ref{maintheoremachievabilityconverse} achieve the optimal rebalancing load by careful construction of the initial and target distributed databases, so as to provide maximal opportunity to perform coded transmissions for rebalancing after the node removal stage. Therefore, we refer to this paradigm of rebalancing schemes as \textit{Coded Data Rebalancing}. We discuss the significance of Theorem \ref{maintheoremachievabilityconverse} through the following observations. 
\begin{itemize}
    \item \textit{\underline{Improvement over uncoded scheme and Optimality:}} Under node removal, since the initial distributed database is $r$-balanced, the number of bits whose replication factor reduces to $r-1$ after node removal is equal to $\lambda N$ (which was amount of data stored in the removed node). 
If naive uncoded transmissions are used to increase the replication factor of these bits to $r$, it is clear that the communication load for rebalancing under node removal will be at least $1$.
Similarly we can show that the rebalancing load for node addition is also at least $1$ under uncoded communications. Thus under uncoded transmissions, the rebalancing load is at least $2$. 
However, our main result shows that the communication load for node removal can be reduced by a multiplicative factor of $r-1$ compared to uncoded schemes, and this is optimal. 
\item \underline{\textit{Structural Invariance:}} We present a class of $r$-balanced databases parametrized by the current number of nodes in the system. Each database in this class corresponding to any particular node cardinality has the same combinatorial structure; we call this property as \textit{structural invariance}. When we initialize the database to be from this class of databases, our presented rebalancing schemes result in another database from the same class after node addition or removal. Thus, this structural invariance is preserved between the initial and target databases. This facilitates the applicability of our rebalancing scheme to any sequence of node addition and node removal rebalancing operations, and also provides seamless indexing of the subfiles across node additions and removals. 
\item \underline{\textit{Optimality of load over sequence of node additions and}} \underline{\textit{ removals:}} Since our database designs and achievability schemes for node removal and addition are optimal and depend only on the replication factor $r$, they are therefore optimal (for a suitably large file size $N$) for a sequence of single node removal and additions also, provided the rebalancing operation takes place after every node removal or addition.
\end{itemize}
}

\begin{example}
\label{main_example}
\textcolor{black}{We illustrate our coded rebalancing schemes through the following example, in which we describe our initial design of the database, then our rebalancing scheme for a node removal, and finally for a node addition. Each rebalancing scheme requires subfile re-indexing to be done, in order to maintain the structural invariance of the database.}

\textcolor{black}{
{\bf Initialization:} Consider \mbox{$K=5$} nodes with replication factor \mbox{$r=3$}. The file $W$ is partitioned into {\mbox{$P(K,K-r)=P(5,2) = 20$}}  subfiles, each indexed by ordered $(K-r) = 2$-sized subsets of $\{1,\hdots,5\}$.  The storage node $i\in[K]$ stores all the subfiles whose indices do not contain $i$.
For instance, storage node $1$ stores $W_{[2~3]}, W_{[3~2]}, W_{[2~4]}, W_{[4~2]}, W_{[2~5]}, W_{[5~2]}$,  $W_{[3~4]},  W_{[4~3]}, W_{[3~5]},  W_{[5~3]}, W_{[4~5]}, W_{[5~4]}$. Thus, each storage node stores \mbox{$P(K-1,K-r)=P(4,2)=12$} subfiles, and thus $\lambda=\frac{12}{20}=\frac{3}{5}$.
}

\textcolor{black}{
{\bf Rebalancing and re-indexing for node removal:}
Consider the case when node 5 is removed. The subfiles in node $5$ now have a smaller replication factor in the surviving nodes. To maintain the replication factor for these subfiles, the contents of node 5 have to be stored among the surviving 4 nodes. In a naive uncoded scheme, the number of transmissions required = 12 subfiles. Using coded rebalancing, we will give a scheme which uses a total of 6 subfile transmissions. 
In our target database, we will have the following.
\begin{align*}
    \text{Node } 1 \text{ stores : } W_{[2]},W_{[3]},W_{[4]}\\
    \text{Node } 2 \text{ stores : } W_{[1]},W_{[3]},W_{[4]}\\
    \text{Node } 3 \text{ stores : } W_{[1]},W_{[2]},W_{[4]}\\
    \text{Node } 4 \text{ stores : } W_{[1]},W_{[2]},W_{[3]},
\end{align*} 
where $W_{[i]}:i\in \{1,2,3,4\}$ refers to a new labelled partition of the file $W$. Note the structure of this target database, where we have indices from $P(K-1,K-1-r)=P(4,1)$. Further, each node $j$ stores the subfiles with indices $[i]$ such that $j\notin [i]$. This target database is essentially of the same structure as the initial database; and our rebalancing scheme realizes this target database in the survivor nodes, thus maintaining the structural invariance. 
}

\textcolor{black}{
In order to do this, the subfiles in node 5 are divided into 4 disjoint groups given as: $\mathcal{G}_4 = \{W_{[1~4]}, W_{[2~4]}, W_{[3~4]}\}, 
\mathcal{G}_3 =  \{W_{[1~3]}, W_{[2~3]}, W_{[4~3]}\}, \mathcal{G}_2 = \{W_{[1~2]}, W_{[3~2]}, W_{[4~2]}\}, \mathcal{G}_1 = \{W_{[2~1]}, W_{[3~1]}, W_{[4~1]}\}$.
In every group, if we consider the set of first elements of each subfile index, we obtain a set of nodes associated to that group. For instance, for the group ${\cal G}_4$, this is $\{1,2,3\}.$ We see that each subfile in the group is available at two of the nodes associated with the group, and missing at exactly one of them (i.e., the first index of that subfile). For instance, $W_{[1~4]}$ is missing at node $1$, while available at nodes $\{2,3\}.$ In our rebalancing scheme, we seek to deliver each such subfile above to one surviving node where it was not present before (indicated by the first index of the subfile), thereby reinstating the replication factor. In the case of ${\cal G}_4$, the subfiles $\{W_{[1~4]}, W_{[2~4]}, W_{[3~4]}\}$ will be delivered to nodes $1,2,3$ respectively. This is done via the data exchange protocol (as in Appendix \ref{app:data_exchange}), which we shall illustrate below. The complete rebalancing scheme consists of running the data exchange protocol for each such group, thus enabling that all the subfiles previously in node $5$ reach their respective target nodes, reinstating the replication factor. }

\textcolor{black}{\underline{Illustrating the data exchange protocol:} We will divide each of the three subfiles in group $\mathcal{G}_4$ into two parts as follows, 
Subfile $W_{[1~4]}$ is split into two equal-sized subfiles $W_{[1~4],2}$ and $W_{[1~4],3}$, $W_{[2~4]}$ is divided into $W_{[2~4],1}$ and $W_{[2~4],3}$, and $W_{[3~4]}$ is divided into $W_{[3~4],1}$ and $W_{[3~4],2}$.
In the protocol for exchanging subfiles within the group $\mathcal{G}_4$, node 1 broadcasts $W_{[2~4],1} \oplus W_{[3~4],1}$, node 2 broadcasts $W_{[1~4],2} \oplus W_{[3~4],2}$, and node 3 broadcasts $W_{[1~4],3} \oplus W_{[2~4],3}$. 
Node 1 can decode $W_{[1~4],2}$ and $W_{[1~4],3}$ from the transmissions of nodes 2 and 3 since it knows the subfiles $W_{[3~4],2}$ and $W_{[2~4],3}$. Node 1 then combines $W_{[1~4],2}$ and $W_{[1~4],3}$ into the subfile $W_{[1~4]}$. It can be seen that the nodes 2 and 3 can also decode $W_{[2~4]}$ and $W_{[3~4]}$, respectively. Thus, the subfiles in group ${\cal G}_4$ have been exchanged among $\{1,2,3\}$ using transmissions of size equal to $\frac{3}{2}^{rds}$ of the size of a subfile.}

\textcolor{black}{
\underline{The complete rebalancing scheme and re-indexing:} Applying the above data exchange protocol to all the $4$ groups independently will reinstate the replication factor for all the subfiles in the removed node $5$, and the total number of subfile transmissions required is $6$. The communication load is therefore $\frac{6}{12}=\frac{1}{2}.$ After the above exchange, each of the $4$ nodes has $15$ subfiles, and thus the new storage fraction $\lambda_{rem}=\frac{3}{4}$.  We will give a method to merge the subfiles and re-index them so that the re-indexing is consistent with our original initialization strategy applied to $4$ nodes with $3$ replicas. For instance, storage node $1$ now has the following subfiles: $W_{[2~3]}, W_{[3~2]}, W_{[2~4]}, W_{[4~2]}$, $W_{[2~5]}, W_{[5~2]},  W_{[3~4]},  W_{[4~3]}$, $W_{[3~5]},  W_{[5~3]}, W_{[4~5]}, W_{[5~4]}, W_{[1~4]}, W_{[1~2]}, W_{[1~3]}$. They are merged and re-indexed at node $1$ as follows:
\textcolor{black}{\begin{eqnarray}
\label{eqnmerging}
W_{[2]} & = & \{W_{[3~2]}, W_{[4~2]}, W_{[2~5]}, W_{[5~2]}, W_{[1~2]}\} \\
W_{[3]} & = & \{W_{[2~3]}, W_{[4~3]}, W_{[3~5]}, W_{[5~3]}, W_{[1~3]}\} \\
W_{[4]} & = & \{W_{[2~4]}, W_{[3~4]}, W_{[4~5]}, W_{[5~4]}, W_{[1~4]}\}
\end{eqnarray}
Such merging and re-indexing is done at each of the nodes. Specifically, the storage at nodes $2,3,4$ in the target database is shown below.
\begin{align*}
    \text{Node } 2 \text{ stores : } W_{[1]},W_{[3]},W_{[4]}\\
    \text{Node } 3 \text{ stores : } W_{[1]},W_{[2]},W_{[4]}\\
    \text{Node } 4 \text{ stores : } W_{[1]},W_{[2]},W_{[3]},
\end{align*} 
where $W_{[1]}$ is formed by merging $\{W_{[2~1]}, W_{[3~1]}, W_{[4~1]}, W_{[5~1]}, W_{[1~5]}\}$ at nodes $\{2,3,4\}$. Similarly $W_{[i]}: i\in\{2,3,4\}$ are  obtained respectively at nodes  apart from $i$, by merging as in (\ref{eqnmerging}). Note the similarity of the storage pattern of the current rebalanced target database with the initial database, illustrating how structural invariance is maintained between the initial and target database. 
}
}

\textcolor{black}{
{\bf Rebalancing and re-indexing for node addition:}  In the case of node addition, there are three steps involved: (i) splitting the subfile (ii) transferring some splitted subfiles to the new node (iii) deleting some splitted subfiles in the original nodes. Consider the case when a new node (labelled node 6) is added. The target database we want to achieve is as follows. 
\begin{itemize}
    \item The subfile indices are chosen from $P(K+1,K+1-r)=P(6,3)$.
    \item Each node $i\in[K]$ stores subfiles with indices not containing $i$. 
\end{itemize}
Thus, once again, we maintain this essential structure of the database. }

\textcolor{black}{
In this case, each subfile is divided (at each node it is stored in) into $6$ parts and re-indexed first. For instance, the subfile $ W_{[2~3]}$ is split as , $ W_{[2~3]} = \{ W_{[1~2~3]},  W_{[4~2~3]},  W_{[5~2~3]}, W_{[6~2~3]}, W_{[2~6~3]}, W_{[2~3~6]} \}$. Note that as $W_{[2~3]}$ is available in the nodes $\{1,4,5\}$, all these parts are also available in those nodes. Among the parts in the above set, $\{W_{[1~2~3]},W_{[4,2,3]},W_{[5,2,3]}\}$ are transferred from nodes $\{1,4,5\}$ (the respective first element of the new indices) respectively to node $6$, and deleted from those specific nodes respectively. For instance, node $1$ transfers $W_{[1~2~3]}$ to node $6$ and deletes it. Repeating this reindex-transmit-delete procedure for each subfile in the database rebalances the database. Clearly, the replication factor is maintained as any deleted parts are stored in the new node first.  Also, we note that half of each original subfile is moved to the new node, and 1-in-6 part of each subfile is removed from the existing nodes. Thus, the new fraction is $\lambda_{add}=1/2$. Further, as the size of transmissions is 10 (original) subfiles, and the communication load is $1$. Finally, it can be seen that the re-indexing is consistent with our original initialization strategy applied to $6$ nodes with $3$ replicas, and hence leads to structural invariance of the database.
}
\end{example}

\subsection{Related work}
Coded transmissions in the presence of local storage have recently been used to greatly reduce the communication load in several multi-receiver communication models, starting from \cite{MaN}. This idea has since then been used in a number of similar scenarios, especially in distributed computing \cite{FundLimitsDistribCom} and distributed data shuffling \cite{speedingup}. The framework for data rebalancing of a distributed database presented in this work enables the abstraction of a communication system with local storage and multiple receivers, and hence permits us to use coded transmissions to reduce the load of communication, similar to \cite{MaN,FundLimitsDistribCom,speedingup}. The results obtained in this work are therefore naturally inspired from those in these works. In particular, the achievability scheme we present is inspired from the scheme in \cite{MaN} which is applicable to a cache-aided noiseless broadcast channel. Our converse proof uses arguments that are similar to those in \cite{FundLimitsDistribCom}. 

There has been a significant amount of work related to designing erasure codes to store data in distributed file systems, which have less storage overhead and also can reconstruct data efficiently in case of node failures. In erasure coding, data is generally divided into blocks. A set of $k$ systematic blocks are used to generate $n-k$ parity blocks, where the parity blocks are linear functions of systematic blocks. The overall set of $n$ blocks comprising of $k$ systematic blocks and $n-k$ parity blocks is referred to as a \textit{stripe}\cite{piggybacking}. Upon node failure, the goal is to reconstruct the failed node using the stripes present in the existing nodes. Each stripe is processed independently and the placement of the reconstructed stripes themselves is not considered specifically.

\textcolor{black}{Further, within the context of distributed storage, there are works which discuss code-conversion \cite{maturana_et_al:LIPIcs:2020:11751} and storage scalability (for instance, \cite{optimalscaling}). The goal in these works is to convert a $n$ node storage system with data encoding using an $(n,k)$ linear block code into another system $n'$ nodes with an $(n',k')$ linear block coded data. The metric to be optimized in \cite{maturana_et_al:LIPIcs:2020:11751} is the number of nodes accessed (which includes node read or written into). The work \cite{optimalscaling} considers only node additions ($n'>n$) and considers minimizing scaling bandwidth which is the total traffic per node involved the scaling process. }

\textcolor{black}{
In this paper, we consider distributed file systems where data is replicated, and we consider data rebalancing operations to correct the data skew caused by both node removal and node additions. We consider the total communication load for rebalancing due to both node removals and additions, and also implicitly consider a sequence of such operations due to a sequence of removals or additions. We show optimal schemes for the same by exploiting careful placement of subfiles of the original data. Our schemes also are structurally invariant in terms of the subfile placement, i.e., the post-rebalancing database structurally mimics the pre-rebalancing state, which enables seamless re-indexing. }



\section{Achievability of Theorem \ref{maintheoremachievabilityconverse}}
\label{achievabilitysection}

\begin{figure*}[t]
\centering
\includegraphics[width=\textwidth]{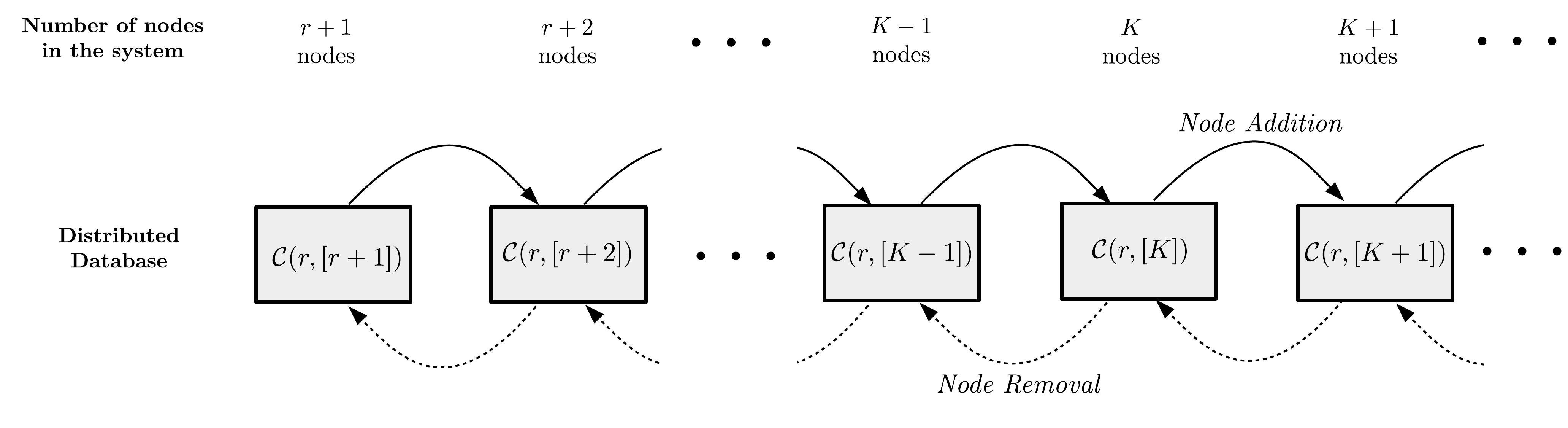}
\caption{The proposed family of $r$-balanced distributed databases $\Cdb(r,[K])$, $K \geq  r+1$.  The databases are parameterized by the number of available storage nodes. Bold arrows and dashed arrows represent rebalancing actions applied for node addition and node removal operations, respectively. The communication loads of rebalancing for any of these node addition and node removal operations are $1$ and $\frac{1}{r-1}$, respectively.}
\label{fig:family_of_databases}
\end{figure*}

In this section we provide construction of $r$-balanced distributed database $\Cdb(r,[K])$ for storing a file across $K$ nodes for any choice of $K$ and $r$ with $2 \leq r \leq K-1$. 
For a given value of $r$, this family of databases is parameterized by the number of nodes $K$. 
We also provide accompanying data rebalancing schemes with rebalancing load $\frac{1}{r-1} + 1$, such that this family of distributed databases is closed under node removal and node addition operations, see Fig~\ref{fig:family_of_databases}.
That is, the node addition operation performed on the database $\Cdb(r,[K])$ yields a target database for $K+1$ nodes that is equivalent to $\Cdb(r,[K+1])$. Similarly any node removal operation on $\Cdb(r,[K])$, irrespective of which one of the $K$ nodes is removed, yields a database that is equivalent to $\Cdb(r,[K-1])$. \textcolor{black}{An example of our construction has been illustrated in Example \ref{main_example}.} 

\subsection{A family of distributed databases}
\label{initialdatabase}
We now describe the proposed family of distributed databases. We require the following notation to describe our construction.
The symbol $P(K,l)$ denotes $K!\,/(K-l)! = K(K-1)\hdots (K-(l-1))$. 
The $m^{\text{th}}$ component of a vector $\p{i}=[i_1~\cdots~i_l]$ will be denoted as $i_m$.
For positive integers $l \leq K$, $\Sc([K],l)$ denotes the set of all vectors $\p{i}$ of length $l$ such that the components of $\p{i}$ belong to the set $[K]$ and all the components of $\p{i}$ are distinct. In other words, $\p{i}=[i_1~i_2~\cdots~i_l] \in \Sc([K],l)$ if and only if $i_1,\dots,i_l$ are distinct and $\{i_1,\dots,i_l\} \subset [K]$. 
For instance, the set $\Sc([3],2)$ consists of the following six vectors
\begin{align*}
[1~2],~[1~3],~[2~1],~[2~3],~[3~1] \text{ and } [3~2].
\end{align*} 
Elementary arithmetic shows that $|\Sc([K],l)| = P(K,l)$. We would like the reader to think of $\p{i} \in \Sc([K],l)$ as a subset of $[K]$ where the order in which the elements of the subset are enumerated in the vector $\p{i}$ matters. For instance, the vectors $[1~2~5], [5~2~1] \in \Sc([5],3)$ arise from two different orderings of the elements of the set $\{1,2,5\}$, and these two vectors must be treated as distinct. 
For any element $k \in [K]$ and any vector $\p{i} \in \Sc([K],l)$, we say that $k$ \emph{belongs to} $\p{i}$, and denote as $k \in \p{i}$, if any one of the components of $\p{i}$ is equal to $k$. 
In case none of the components of $\p{i}$ is equal to $k$, we say that $k$ \emph{does not belong to} $\p{i}$ and denote this as $k \notin \p{i}$. 
 
\emph{The proposed distributed database scheme $\Cdb(r,[K])$:} 
Our $r$-balanced distributed database for $K$ storage nodes is constructed as follows. As in the statement of Theorem \ref{maintheoremachievabilityconverse} we assume that the size $N$ of the file $W$ is divisible by $(r-1)P(K+1,K+1-r)$. 
We partition the given file $W$ into $P(K,K-r)$ subfiles and index the subfiles using the vectors in $\Sc([K],K-r)$.
The subfiles of $W$ are denoted as $W_{\p{i}}$, $\p{i} \in \Sc([K],K-r)$. 
We assume that the length of each of the subfiles is same, i.e., each subfile is of size $N/P(K,K-r)$~bits. The contents of the storage nodes are given by the following design
\begin{itemize}
\item the distributed database $\Cdb(r,[K])$ stores the subfile $W_{\p{i}}$ in storage node $k$ if and only if $k \notin \p{i}$.
\end{itemize}

Since the length of the vector $\p{i} \in \Sc([K],K-r)$ is $K-r$, there exist exactly $r$ elements in $[K]$ that do not belong to $\p{i}$. The corresponding storage nodes, i.e., those with indices $k \notin \p{i}$, store $W_{\p{i}}$, and the other nodes do not store this subfile. Thus the replication factor of every subfile is precisely $r$.

The number of subfiles stored in node $k$ is equal to the size of the set $\{\,\p{i} \in \Sc([K],K-r)~|~k \notin \p{i}\,\}$, which is equal to $(K-1)(K-2)\cdots r = P(K-1,K-r)$. Thus, the ratio of the number of bits stored in node $k$ to the size of $W$ is 
\begin{equation*}
\lambda = \frac{P(K-1,K-r)}{P(K,K-r)}= \frac{(K-1)(K-2)\cdots\,r}{K(K-1)\cdots\,(r+1)} =\frac{r}{K}.
\end{equation*} 
We conclude that $\Cdb(r,[K])$ is an $r$-balanced distributed database for $K$ nodes.

\begin{example}
Consider \mbox{$K=4$} nodes with replication factor \mbox{$r=2$}. The file $W$ is partitioned into \mbox{$P(K,K-r)=P(4,2) = 12$} subfiles, each indexed by a length $2$ vector in $\Sc([4],2)$. Each storage node stores \mbox{$P(K-1,K-r)=P(3,2)=6$} subfiles.
For instance, storage node $1$ stores $W_{[2~3]}, W_{[2~4]}, W_{[3~2]}, W_{[3~4]}, W_{[4~2]}, W_{[4~3]}$, and node $2$ stores $W_{[1~3]}, W_{[1~4]}, W_{[3~1]}, W_{[3~4]}, W_{[4~1]}, W_{[4~3]}$.
\end{example}
\textcolor{black}{\begin{remark}
\underline{Relationship to the Ali-Niesen scheme \cite{MaN}:} The choice of the family of distributed databases is closely related to the placement phase of the Ali-Niesen coded caching scheme \cite{MaN}, the careful reader will note. While in the Ali-Niesen scheme the subfile indices are indicative of the users in which a subfile is present, here we use the collection of users where a subfile is absent as the subfile indexing. This particular technique of `symmetric' placement enables us to create maximal coding opportunities during the rebalancing phase, as well as ensure that structural invariance of the database can be maintained after rebalancing due to node removal or addition. There is a distinction with the placement scheme in \cite{MaN} however, which is the ordering on the subfile index. This ordering enables us to decide the target nodes for subfiles during the rebalancing phase after node removal or addition. This incurs a cost in the size of the file $N$ compared to \cite{MaN} (which uses coding of subfiles to achieve a different end). 
\end{remark}
With the initial database ${\cal C}(r,[K])$ in place, we now give the rebalancing scheme for node addition and for node removal. The node removal scheme exploits the existence of replicated subfiles placed in a structured manner in the original database for doing coded transmissions, and thus reducing the rate. The node addition scheme uses for its description the combinatorial structure of the initial database. In both the scenarios, we achieve our goal of least communication load and also structural invariance of the target rebalanced databases compared to the initial database. The two schemes are in some sense counterparts to each other, with the node addition scheme working logically similar to the node removal scheme, but with the steps run in reverse. Since the node addition scheme is easier to describe, we begin with that in Section \ref{rebalancingschemeaddition}, and subsequently discuss the node removal scheme in Section \ref{rebalancingschemeremoval}. }
\subsection{Data Rebalancing for Node Addition}
\label{rebalancingschemeaddition}
We consider the scenario where $K$ nodes are storing a file $W$ with replication factor $r$ using the distributed database scheme $\Cdb(r,[K])$. 
A new node, denoted using the index $K+1$, is introduced into the system. This new node does not contain any information. 
We will now provide an algorithm to allow the $K$ pre-existing nodes to communicate with the new node in order to establish an $r$-balanced distributed database that stores the file $W$ across all $K+1$ nodes.

The proposed algorithm involves the partitioning of each subfile $W_{\p{i}}$ present in $\Cdb(r,[K])$ into $K+1$ parts, and providing these parts with new indexing labels. 
Note that the length of $\p{i}$ is $K-r$, and there exist $r$ elements in $[K]$, say, $j_1,\dots,j_r$ that do not belong to $\p{i}$.
Assume that $j_1 < j_2 < \cdots < j_r$. Also note that $K+1 \notin \p{i}$, since $\p{i} \in \Sc([K],K-r)$. 
Let $\p{i}=[i_1~i_2~\cdots~i_{K-r}]$.
In order to partition a subfile $W_{\p{i}}$, we split the contents of the subfile into $K+1$ equal-sized parts, and label these parts with the following length \mbox{$K-r+1$} vectors (in that order) 
\begin{align*}
&[j_1~i_1~\cdots~i_{K-r}],~[j_2~i_1~\cdots~i_{K-r}],~\cdots~,~[j_r~i_1~\cdots~i_{K-r}], \\
&[K+1~i_1~\cdots~i_{K-r}],~[i_1~K+1~i_2~\cdots~i_{K-r}],~\cdots,\\
&[i_1~i_2~\cdots~i_{K-r}~K+1].
\end{align*} 
The first $r$ of the above vectors are obtained by prefixing $\p{i}$ with $j_1,\dots,j_r$, respectively. The remaining $K-r+1$ vectors are obtained by all possible insertions of the component $K+1$ into the vector $\p{i}$. 
These new parts are denoted by $W_{[j_1~i_1~\cdots~i_{K-r}]}, \dots, W_{[i_1~i_2~\cdots~i_{K-r}~K+1]}$, respectively.
Note that the new indices are vectors from the set $\Sc([K+1],K-r+1)$.

The size of each of these $K+1$ new subfiles is 
\begin{equation} \label{eq:size_of_parts_node_addition}
\frac{N}{P(K,K-r)} \times \frac{1}{K+1} = \frac{N}{P(K+1,K+1-r)} \text{ bits}.
\end{equation} 

The data rebalancing scheme for node addition is as follows. 
For each $k \in [K]$, the node $k$ partitions each of the subfiles available to it into $K+1$ parts using the technique described in the previous paragraph. 
The subfiles available at node $k$ are $W_{\p{i}}$, $\p{i} \in \Sc([K],K-r)$ and $k \notin \p{i}$.
Note that when $W_{\p{i}}$ is partitioned by node $k$, one of the resulting parts will have the index $[k~\p{i}]$. After partitioning, node $k$ transfers the following parts to the new node 
\begin{equation*}
W_{[k~\p{i}]}, \text{ where } \p{i} \in \Sc([K],K-r) \text{ and } k \notin \p{i},
\end{equation*}
while removing them from its own memory. 
The rest of the new parts are stored in node $k$.
Node $K+1$ receives all such transmissions from each of the $K$ pre-existing nodes, and stores the received contents in its memory. 
It is not difficult to observe that the parts remaining in node $k$ correspond to the set of indices 
$\{\p{i}' \in \Sc([K+1],K-r+1) ~|~ k \notin \p{i}'\}$,
and the parts now stored in node $K+1$ have the indices in the set
$\{\p{i}' \in \Sc([K+1],K-r+1) ~|~ K+1 \notin \p{i}'\}$.
This placement of contents of $W$ across $K+1$ nodes is identical to the distributed database scheme $\Cdb(r,[K+1])$.

The number of parts communicated by node $k$ to node $K+1$ is $P(K-1,K-r)$, and the number of parts remaining with node $k$ is $K \times P(K-1,K-r) = P(K,K-r+1)$.
Using~\eqref{eq:size_of_parts_node_addition}, the fraction of the overall file stored in each of the $K+1$ nodes after data rebalancing for node addition is
\begin{equation*}
\lambda_{add} = \frac{P(K,K-r+1)}{P(K+1,K-r+1)} = \frac{r}{K+1} = \lambda\frac{K}{K+1},
\end{equation*} 
which is as required for an $r$-balanced scheme.
The total number of bits communicated during rebalancing is 
\begin{equation*}
K \times \frac{P(K-1,K-r)\,N}{P(K+1,K-r+1)} = \frac{rN}{(K+1)} = \lambda_{add}N.
\end{equation*}
Thus, the communication load of this rebalancing scheme for node addition is $L_{add}=1$.

\subsection{Coded Data Rebalancing for Node Removal} 
\label{rebalancingschemeremoval}
We now provide a rebalancing scheme for mitigating data skew when one of the nodes in the $r$-balanced distributed database $\Cdb(r,[K])$ fails. The subfiles that were originally available in the failed node are now replicated only $(r-1)$ times in the surviving nodes. The main objective of our data balancing scheme is to place a new copy of any such subfile $W_{\p{i}}$ in the node with index $i_1$, where $i_1$ is the first component of $\p{i}$. This will ensure that all subfiles are replicated $r$ times across $(K-1)$ nodes. The subfiles are then combined in a specific way and re-indexed so that the resulting database is structurally identical to $\Cdb(r,[K-1])$.

\subsubsection{Review of a Data Exchange Protocol} \label{sec:sub:data_exchange} 

As one of the components of our data rebalancing scheme, we make use of a communication efficient protocol for exchanging data between a group of $r$ storage nodes. 
This protocol is used when $r$ nodes are connected by a common broadcast link, each node stores a distinct $(r-1)$ subset of a set of $r$ files $B_1,\dots,B_r$, and each node demands the unique file that is not available in its memory.
If the size of each of the files $B_1,\dots,B_r$ is $\ell$ bits, the overall communication cost of this protocol, i.e., the total number of bits broadcast by all the $r$ nodes, is $\ell\, r/(r-1)$ bits.
This protocol is known in the literature; for instance,~\cite{CodedMapreduce} uses this for coded MapReduce.
For the sake of completeness, we provide a brief description of this protocol in Appendix~\ref{app:data_exchange}.

\subsubsection{Data Rebalancing}

Assume that an arbitrary node $k$ is removed from the distributed database $\Cdb(r,[K])$. We denote the index set of the remaining nodes as $\Kt=[K]\setminus  k$.
Since node $k$ is removed and since $\Cdb(r,[K])$ is an $r$-balanced scheme, the subfiles that were not originally stored in node $k$ are still replicated $r$ times among the surviving nodes.
However, each of the subfiles originally available in node $k$ is now available at only $(r-1)$ of the remaining $(K-1)$ nodes. Let $\At_k$ denote the index set of these subfiles, i.e., $\At_k = \{ \p{i} \in \Sc([K],K-r)~|~ k \notin \p{i} \}$. 

We partition the set of subfiles $\{W_{\p{i}}~|~\p{i} \in \At_k\}$, into groups, each of which will be coded and communicated together for data rebalancing.
The groups are indexed by length $(K-1-r)$ vectors $\p{i'} \in \Sc([K],K-1-r)$ where $k \notin \p{i'}$, i.e., the vectors $\p{i'} \in \Sc(\Kt,K-1-r)$. 
For each such $\p{i'}$, we define
\begin{equation*}
\Gt_{\p{i'}} = \{ \p{i} \in \At_k ~|~ [i_2~i_3~\cdots~i_{K-r}] = \p{i'} \}.
\end{equation*}
The number of vectors $\p{i}$ such that $[i_2~\cdots~i_{K-r}] = \p{i'}$ and $\p{i} \in \At_k$, i.e., the number of choices for the component $i_1$ such that $i_1 \notin \p{i'}$ and $i_1 \neq k$, is $r$. We also observe that for any two distinct $\p{i'}, \p{j'} \in \Sc(\Kt,K-1-r)$, the sets $\Gt_{\p{i'}}$ and $\Gt_{\p{j'}}$ are non-intersecting. Thus, these $r$-sized groups form a partition of $\At_k$.
Since $|\At_k| = P(K-1,K-r)$, we conclude that the number of such groups is $P(K-1,K-r)/r$.

The objective of our data rebalancing scheme is to replicate $W_{\p{i}}$, $\p{i} \in \At_k$, at the storage node with index $i_1$. Note that since $i_1 \in \p{i}$, this subfile was not originally present in the node $i_1$.
We achieve this objective by running one round of data exchange protocol for each of the $P(K-1,K-r)/r$ groups of subfiles, resulting in as many rounds of the protocol.
Now, consider the subfiles $W_{\p{i}}$, $\p{i} \in \Gt_{\p{i'}}$, belonging to one of these groups. Let $p_1,\dots,p_r$ be such that
$\Gt_{\p{i'}} = \{[p_1~\p{i'}],\,\cdots,\,[p_r~\p{i'}]\}$.
Using the facts that $p_1,\dots,p_r \notin \p{i'}$ and $p_1,\dots,p_r \neq k$, we observe that the subfile $W_{[p_1~\p{i'}]}$ is available at the nodes $p_2,\dots,p_r$, and we desire to replicate this subfile at node $p_1$.
In all, each of the $r$ subfiles $W_{[p_1~\p{i'}]},\dots,W_{[p_r~\p{i'}]}$ is available in a unique $(r-1)$-sized subset of the $r$ storage nodes $p_1,\dots,p_r$, and is required to be replicated at the remaining node as well.
We can achieve the replication of the subfiles $W_{[p_1~\p{i'}]},\dots,W_{[p_r~\p{i'}]}$ at nodes $p_1,\dots,p_r$, respectively, using the data exchange protocol~\cite{CodedMapreduce} referred to in Section~\ref{sec:sub:data_exchange}. 
Since the size of each of the $r$ subfiles is $\ell = \frac{N}{P(K,K-r)}$~bits, the communication cost of the protocol is
\begin{equation*}
\frac{\ell\,r}{(r-1)} = \frac{N\,r}{(r-1)\,P(K,K-r)}~\text{ bits}.
\end{equation*} 

The above data exchange is performed for each group $\Gt_{\p{i'}}$, $\p{i'} \in \Sc(\Kt,K-1-r)$. Since the number of groups is $P(K-1,K-r)/r$, the overall communication cost of our rebalancing scheme is
\begin{equation*}
\frac{N\, P(K-1,K-r)}{(r-1)\,P(K,K-r)}  = \frac{N\, r}{(r-1) \, K} = \frac{N \lambda}{(r-1)},
\end{equation*}  
and the resulting communication load is $L_{rem}=1/(r-1)$.
     
We now analyze the memory utilization at the surviving nodes at the end of rebalancing operation. Each node $m \neq k$, has been originally storing subfiles with indices in the set $\At_m = \{ \p{i}~|~m \notin \p{i} \}$, and will additionally store the subfiles corresponding to the indices $\{ \p{i} \in \At_k~|~i_1 = m \}$. Thus, the number of subfiles in node $m$ after rebalancing is the sum of the sizes of these sets, which is
\begin{equation*}
P(K-1,K-r) + P(K-2,K-1-r) \! = \! K\, P(K-2,K-1-r).
\end{equation*} 
Multiplying this by the size of each subfile, we obtain the overall size of the contents of node $m$ after rebalancing
\begin{align*}
K\, P(K-2,K-1-r) \times \frac{N}{P(K,K-r)} = \frac{Nr}{K-1}.
\end{align*}  
Thus, $\lambda_{rem} = \frac{r}{(K-1)} = \lambda\frac{K}{(K-1)}$.

Our rebalancing scheme replicates each $W_{\p{i}}$, $\p{i} \in \At_k$, at exactly one of the surviving nodes $\Kt$, increasing the replication factor of these subfiles among the nodes in $\Kt$ from $(r-1)$ to $r$. 
The subfiles which were not contained in node $k$ in $\Cdb(r,[K])$, already have a replication factor of $r$ among the nodes in $\Kt$. Thus, we conclude that the achieved target database is an $r$-balanced database across $K-1$ nodes. 

\subsubsection{Re-indexing and Structural Invariance}

We now combine the subfiles ($K$ of them at a time) available in the nodes $\Kt$ after rebalancing, and then re-index them. 
This re-indexing operation uses vectors from $\Sc(\Kt,K-1-r)$, i.e., vectors $\p{i'}$ of length $K-1-r$ whose components are distinct elements of $\Kt = [K] \setminus  k$.
Our objective is to re-index the subfiles such that each node \mbox{$m \in \Kt$} consists of all the re-indexed subfiles whose indices $\p{i'} \in \Sc(\Kt,K-1-r)$ satisfy $m \notin \p{i'}$.
Since $|\Kt|=K-1$, this ensures that the new database is identical to $\Cdb(r,[K-1])$, up to a relabelling of the storage nodes.

Consider any $\p{i'} \in \Sc(\Kt,K-1-r)$. There exist distinct $j_1,\dots,j_r \in \Kt$ such that $j_1,\dots,j_r \notin \p{i'}$. Further, $k \notin \p{i'}$. Assuming $j_1 < j_2 < \cdots < j_r$, a new re-indexed subfile $W_{\p{i'}}$ is obtained by concatenating the contents of the following $K$ original subfiles (in that order) whose indices are
\begin{align*}
&[j_1~\p{i'}],~[j_2~\p{i'}],~\cdots~,~[j_r~\p{i'}], \\
&[k~i'_1~\cdots~i'_{K-1-r}],~[i'_1~k~i'_2~\cdots~i'_{K-1-r}],~\cdots,\\
&[i'_1~i'_2~\cdots~i'_{K-1-r}~k].
\end{align*}
Note that, after rebalancing, any node $m \notin \p{i'}$, $m \neq k$, stores all the above $K$ subfiles. While rebalancing delivers $W_{[m~ \p{i'}]}$ to node $m$, the other $(K-1)$ subfiles are already present in this node by the design of $\Cdb(r,[K])$.  
Thus, for every $\p{i'} \in \Sc(\Kt,K-1-r)$ and every choice of $m \notin \p{i'}$, $m \in \Kt$, node $m$ can perform this re-indexing operation and store the re-indexed file $W_{\p{i'}}$ in its memory.
It is straightforward to see that $W_{\p{i'}}$, $\p{i'} \in \Sc(\Kt,K-1-r)$, form a partition of the file $W$, and that node $m$ stores $W_{\p{i'}}$ if and only if $m \notin \p{i'}$.
Thus the rebalanced database is identical to $\Cdb(r,[K-1])$. 

\subsection{Rebalancing Load}
The communication load of our rebalancing scheme for the removal of any node $k$ is equal to $L_{rem}=1/(r-1)$, and communication load for node addition is $L_{add}=1$. We conclude that the rebalancing load for our scheme is $\frac{1}{(r-1)} + 1$.


\section{Converse of Theorem \ref{maintheoremachievabilityconverse}}
\label{sectionconverse}
We first consider the node addition case. Noting the fact that the new node arrives without any stored information, it is clear that any rebalancing scheme for node addition must necessarily involve communicating $\lambda_{add}N$ bits to the new node. Hence $L_{add}({\cal R}'_{{\cal C},{\cal C}'})\geq 1$ for any rebalancing scheme ${\cal R}'$ and any initial and target databases ${\cal C},{\cal C}'.$ 

We now obtain the converse for the case when there is one failed node (the node $K$, without loss of generality) among the nodes $[K]$. The proof of the converse proceeds quite similar to the proof of converse in \cite{FundLimitsDistribCom} in the context of distributed computing with coded data shuffling (Section VI in \cite{FundLimitsDistribCom}).

We assume that the file $W$ is a uniform random variable taking values from ${\mathbb F}_2^N$. For $k\in[K]$, we recall that $C_k$ denotes the set of all bits of $W$ which are available in the storage of node $k$ in the initial database. For a subset $S\subseteq [K]$, let $C_S=\bigcup\limits_{k\in S} C_k$. 

For a subset of bits $B\subseteq C_K$, a subset of the nodes $S\subseteq [K-1]$, and some $j\in\{0,1,\hdots,|S|\}$, let $a_{B}^{j,S}$ denote the number of distinct bits of $B$ which are available in exactly $j$ of the nodes in $S$, and not anywhere else, i.e.,
$$a_{B}^{j,S}=\sum_{{\cal J}\in\binom{S}{j}}\bigl\vert \left(\left(\bigcap\limits_{k\in {\cal J}}C_k\right)\cap B\right)\backslash \left(\bigcup\limits_{k\in[K-1]\backslash{\cal J}} C_k\right)\bigr\vert,$$ 
where $\binom{S}{j}$ denotes the set of $j$-sized subsets of $S$.

Based on our assumptions regarding the system before and after the node failure, we have the following statements to be true. 
\begin{align}
\label{eqn204}
\sum_{j=1}^{K-1}a_{C_K}^{j,[K-1]}&=|C_K|=\lambda N. \\
\label{eqn205}
\sum_{j=1}^{K-1} ja_{\{c\}}^{j,[K-1]}&=r-1,~\forall c\in C_K.\\
\label{eqn206}
\sum_{j=1}^{K-1} ja_{C_K}^{j,[K-1]}&=(r-1)\lambda N
\end{align}

Equation (\ref{eqn204}) holds because we assume $r\geq 2$ (otherwise rebalancing after node $K$ failure would be impossible). Also, (\ref{eqn205})  is true since exactly one repetition of bit $c\in C_K$ is unavailable after the failure of node $K$, and (\ref{eqn205}) leads to (\ref{eqn206}). 

 After the failure of node $K$, the surviving part of the database has replication factor $r-1$ for bits in $C_K$. We then want to design the rebalancing scheme such that target database has replication factor $r$ for bits of $C_K$ also. This means any rebalancing scheme should be designed so that each  bit $c\in C_K$ is to be communicated to at least one node in $[K-1]$ which does not already contain $c$. Recalling the fact that $2\leq r\leq (K-1)$ by Remark \ref{remark_r_bounds_K}, we also note that we should have $K\geq 3.$ We now formalize the aspects of any valid rebalancing scheme now.
  
For $k \in [K-1]$, we recall that $C_k'$ denotes the set of bits stored in node $k$ in the target database. Let $D'_k\triangleq C_k'\backslash C_k$, denote the set of `new' bits to be stored in the node $k\in[K-1]$ respectively, in the target database. Further define 
\begin{align}
\label{eqndkdefinition} 
D_k\triangleq (D_k'\cap C_K)\backslash \left(\cup_{k_1 < k }D_k\right)
\end{align}
In other words, $D_k$ denotes the set of bits of $C_K$ to be stored in node $k$, which has not already been stored in any nodes in the set $\{1,\hdots,k-1\}$.

For a subset $S\subseteq [K-1]$, we also denote $D_S\triangleq \cup_{k\in S}D_k.$ We then have the following.  
\begin{align}
\label{eqna1}
    \bigcup\limits_{k\in[K-1]}D_k&=D_{[K-1]}=C_K\\
    \label{eqna2}
    D_k\cap D_{k'}&=\phi, \forall ~\text{distinct}~k,k'\in[K-1], \\
    \label{eqna3}
    D_k\cap C_k&=\phi, \forall ~k\in[K-1]. 
\end{align}
Equation (\ref{eqna1}) is true because all the bits of $C_K$ have to necessarily be stored in at least one surviving node. Equations (\ref{eqna2},\ref{eqna3}) follow from (\ref{eqndkdefinition}).

For $k\in [K-1]$, let $X_k$ denote the set of transmitted bits by node $k$ to perform the rebalancing. Note that since the messages to be exchanged are subsets of $C_K$, we thus have that $H(X_k|C_k)=0$. For a subset $S\subseteq [K-1]$, we denote $X_S\triangleq \{X_k:k\in S\}$. We then want, 
\textcolor{black}{$$H(D_k|X_{[K-1]\backslash k},C_k)=0.$$} For a subset $S\in[K-1]$, we define the quantity $Y_S$ as follows.
$$Y_S=\{D_k:k\in S\}\cup\{C_k:k\in S\}.$$

Following the technique in \cite{FundLimitsDistribCom}, we first prove a lower bound on the quantity $H(X_S|Y_{\overline{S}})$, where $\overline{S}=[K-1]\backslash S$. The converse will then follow by substituting $S=[K-1]$. The lower bound on $H(X_S|Y_{\overline{S}})$ is given by the following lemma.
\begin{lemma}
\label{lowerboundsubset}
For $S\subseteq [K-1]$ such that $|S|\geq 2$, we have $$H(X_S|Y_{\overline{S}})\geq \sum_{j=1}^{|S|-1}\frac{a_{D_S}^{j,S}}{j}.$$
\end{lemma}
\begin{IEEEproof}
We prove the lemma by induction. First consider the base case when $|S|=2$, and without loss of generality let $S=\{1,2\}$. Then $a_{D_1}^{1,\{1,2\}}$, by definition and by (\ref{eqna3}), must lie in node $2$ only and nowhere else in the $K-1$ nodes. By a similar argument for $a_{D_2}^{1,\{1,2\}}$, we must have that $H(X_S|Y_{\overline{S}})\geq a_{D_1}^{1,\{1,2\}}+a_{D_2}^{1,\{1,2\}}=a_{D_{\{1,2\}}}^{1,\{1,2\}},$ which proves the base case.

Now we assume that the statement holds for all subsets of $[K-1]$ of size $t=|S|-1$.  We then want to show it for $S$. We have the following expressions.
\begin{align*}
H(X_S|Y_{\overline{S}})&=\frac{1}{|S|}\sum_{k\in S}H(X_S,X_k|Y_{\overline{S}})\\
&=\frac{1}{|S|}\sum_{k\in S} (H(X_S|X_k,Y_{\overline{S}})+H(X_k|Y_{\overline{S}}))\\
&\textcolor{black}{\geq}  \frac{1}{|S|}\left(\sum_{k\in S} H(X_S|X_k,Y_{\overline{S}})+H(X_S|Y_{\overline{S}})\right).
\end{align*}
By re-arranging the terms, we get
\begin{align}
\nonumber
H(X_S|Y_{\overline{S}})&\textcolor{black}{\geq} \frac{1}{t}\sum_{k\in S} H(X_S|X_k,Y_{\overline{S}}) \\
\nonumber
&\geq \frac{1}{t}\sum_{k\in S} H(X_S|X_k,C_k,Y_{\overline{S}})\\
\label{eqn301}
H(X_S|Y_{\overline{S}}) &\geq \frac{1}{t}\sum_{k\in S} H(X_S|C_k,Y_{\overline{S}}),
\end{align}
where (\ref{eqn301}) follows as $H(X_k|C_k)=0$.
Now, we have for any $k\in S$, 
\begin{align}
\nonumber
H&(X_S,D_k|C_k,Y_{\overline{S}})=H(X_S|C_k,Y_{\overline{S}})+H(D_k|X_S,C_k,Y_{\overline{S}})\\
\label{eqn302}
 &=H(X_S|C_k,Y_{\overline{S}})=H(D_k|C_k,Y_{\overline{S}})+H(X_S|D_k,C_k,Y_{\overline{S}}),
\end{align}
where the second equality in (\ref{eqn302}) follows because $D_k$ is decodable given $X_S,Y_{\overline{S}},$ \textcolor{black}{and $C_k$}. 

We now reduce the two components of the last expression in (\ref{eqn302}) separately. Firstly, because $D_k$ and $D_{k'}$ are independent (as the $D_k$s form a partition of $C_K$ by (\ref{eqna1}),(\ref{eqna2})), we also have 
\begin{align}
\label{eqn401}
H(D_k,D_{\overline{S}}|C_k,C_{\overline{S}})=H(D_k|C_k,C_{\overline{S}})+H(D_{\overline{S}}|C_k,C_{\overline{S}})
\end{align}
We now have the following expressions.
\begin{align}
\nonumber
H(D_k|C_k,Y_{\overline{S}})&=H(D_k|C_k,D_{\overline{S}},C_{\overline{S}})\\
\label{eqn501}
&=H(D_k|C_k,C_{\overline{S}}),\\
\label{eqn502}
&=H(D_k|C_{\overline{S}\cup\{k\}}),
\end{align}
 where (\ref{eqn501}) follows from (\ref{eqn401}). The expression in (\ref{eqn502}) is the number of bits of $D_k$ which are present only in $S\backslash k$ (since every bit of $D_k$ must be present in at least one of the $K-1$ surviving nodes). Thus we have from (\ref{eqn502}), 
 \begin{align}
 \label{eqn601}
H(D_k|C_k,Y_{\overline{S}}) =H(D_k|C_{\overline{S}\cup\{k\}})=\sum_{j=1}^t a_{D_k}^{j,S\backslash k}.
 \end{align}
 We now bound the second term of the last expression of (\ref{eqn302}). We have the following.
 \begin{align}
 \nonumber	
H(X_S|D_k,C_k,Y_{\overline{S}})&= H(X_S|Y_{\overline{S}\cup\{k\}})\\
&=H(X_{S\backslash k}|Y_{\overline{S}\cup\{k\}})\\
 \label{eqn701}
& \geq \sum_{j=1}^{t-1}\frac{a_{D_{S\backslash  k}}^{j,S\backslash  k}}{j},
 \end{align} 
 where the second equality follows because $Y_{\overline{S}\cup k}$ contains $C_k$ and $H(X_k|C_k)=0$, and the  last inequality follows by the induction hypothesis. 
 Now, by using (\ref{eqn701}) and (\ref{eqn601}) in (\ref{eqn302}), we get
 \begin{align}
 \label{eqn703}
 H(X_S,D_k|C_k,Y_{\overline{S}})\geq  \sum_{j=1}^t a_{D_k}^{j,S\backslash k} +\sum_{j=1}^{t-1}\frac{a_{D_{S\backslash  k}}^{j,S\backslash  k}}{j}.
 \end{align}
 Now, 
 \begin{align}
 \nonumber
& \sum_{k\in S} a_{D_k}^{j,S\backslash k}\\
\nonumber
&\small = \sum_{k\in S} \sum_{n=1}^ {|D_k|}{\mathbb I}(n^{th}\text{bit of}~D_k~\text{is stored nowhere except}~j~\text{nodes of}~S\backslash k)\\
\nonumber&\small = \sum_{k\in S} \sum_{n=1}^ {|D_k|}{\mathbb I}(n^{th}\text{bit of}~D_k~\text{is stored  nowhere except}~j~\text{nodes of}~S)\\
\label{eqn801}
&=\sum_{k\in S}a_{D_k}^{j,S}=a_{D_S}^{j,S},
 \end{align}
 where the second equality holds because no bits of $D_k$ are in $C_k$. We have also the following,
 \begin{align}
 \nonumber
 \sum_{k\in S}a_{D_{S\backslash  k}}^{j,S\backslash  k}&=\sum_{k\in S}(a_{D_{S}}^{j,S\backslash  k}-a_{D_k}^{j,S\backslash  k})\\
 \label{eqn901}&=\sum_{k\in S}a_{D_{S}}^{j,S\backslash  k}-a_{D_S}^{j,S},
 \end{align}
 where the second equality follows from (\ref{eqn801}).  Further, 
 \begin{align}
 \nonumber
 &\sum_{k\in S}a_{D_{S}}^{j,S\backslash  k} \\
  \nonumber &\small = \sum_{k\in S} \sum_{n=1}^ {|D_S|}{\mathbb I}(n^{th}\text{bit of}~D_S~\text{is stored nowhere except}~j~\text{nodes of}~S\backslash k)\\
 \nonumber&\small = \sum_{k\in S} \sum_{n=1}^ {|D_S|}\left({\mathbb I}(n^{th}\text{bit of}~D_S~\text{is stored nowhere except}~j~\text{nodes of}~S)\right.\\
 \nonumber&\small~~~~~~~~~~~~~~~~~\times\left.{\mathbb I}(n^{th}~\text{bit of}~D_S~\text{is not stored in node}~k)\right)\\
 \nonumber&\small = \sum_{n=1}^ {|D_S|}\left({\mathbb I}(n^{th}\text{bit of}~D_S~\text{is stored nowhere except}~j~\text{nodes of}~S)\right.\\
  \nonumber
&\small ~~~~~~~~~~~~~~~~~\times \sum_{k\in S}\left.{\mathbb I}(n^{th}~\text{bit of}~D_S~\text{is not  stored in node}~k)\right)\\
\label{eqn902}
&=a_{D_S}^{j,S}(|S|-j).
 \end{align}
 where the last equality is true because \begin{align*}\textcolor{black}{\sum_{n=1}^ {|D_S|}}{\mathbb I}(n^{th}&\text{bit of}~D_S~\text{is  stored nowhere except}~j~\text{nodes of}~S)\\&=a_{D_S}^{j,S}.
 \end{align*}
 Summing over all $k\in S$ on both sides of (\ref{eqn703}), and using (\ref{eqn801}) (\ref{eqn901}) and (\ref{eqn902}), we get
\begin{align} 
\nonumber
\sum_{k\in S}H(X_S,D_k|C_k,Y_{\overline{S}})&\geq \sum_{j=1}^t a_{D_S}^{j,S}+\sum_{j=1}^{t-1}\frac{a_{D_S}^{j,S}(|S|-1-j)}{j}\\
\label{eqn903}
 &\geq a_{D_S}^{t,S}+\sum_{j=1}^{t-1}\frac{t.a_{D_S}^{j,S}}{j}.
 \end{align}
 By using (\ref{eqn903})  and the second equality of (\ref{eqn302}) in (\ref{eqn301}), we get 
 \begin{align}
 H(X_S|Y_{\overline{S}})\geq \frac{a_{D_S}^{t,S}}{t}+\sum_{j=1}^{t-1}\frac{a_{D_S}^{j,S}}{j}=\sum_{j=1}^{t}\frac{a_{D_S}^{j,S}}{j}.
 \end{align} 
 This completes the proof of the lemma. 
 \end{IEEEproof}
 By applying Lemma \ref{lowerboundsubset} to the set $S=[K-1]$ and noting that $a_{D_{[K-1]}}^{j,[K-1]}=a_{C_K}^{j,[K-1]}$, we have
 $$
 \textcolor{black}{H(X_{[K-1]})\geq \sum_{j=1}^{K-2}\frac{a_{C_K}^{j,[K-1]}}{j}=\sum_{j=1}^{K-1}\frac{a_{C_K}^{j,[K-1]}}{j},}
 $$
 where the last equality holds as $a_{C_K}^{K-1,[K-1]}=0$ \textcolor{black}{since $r\leq K-1$ by Remark \ref{remark_r_bounds_K}}. As $\frac{1}{j}$ is convex in $j$ and since $\sum_{j=1}^{K-1}a_{C_K}^{j,[K-1]}=\lambda N$ by (\ref{eqn204}), we thus have from the above last equation
\begin{align*}
 H(X_{[K-1]})&\geq \lambda N\sum_{j=1}^{K-1}\frac{a_{C_K}^{j,[K-1]}}{\lambda N}.\frac{1}{j}\\
 &\geq \frac{\lambda N}{\sum_{j=1}^{K-1}\frac{ja_{C_K}^{j,[K-1]}}{\lambda N}}=\frac{\lambda N}{r-1},
\end{align*}
where the last expression is true by (\ref{eqn206}). The converse for the rebalancing load under node removal is then complete by the definition of the load in this case. 

By definition of the optimal load $L^*(r)$, we have therefore showed that the lower bound expression in Theorem \ref{maintheoremachievabilityconverse} is true. This completes the converse proof. 


\appendices

\section{Review of Data Exchange Protocol} \label{app:data_exchange}

Without loss of generality, label the $r$ nodes as $1,\dots,r$, respectively, and assume that each node $m \in [r]$ contains the files $B_j$, \mbox{$j \in [r] \setminus \{m\}$}. That is, the only file not available at node $m$ is $B_m$. The objective of the data exchange protocol is to deliver the file $B_m$ to node $m$ for each $m \in [r]$.

We split each file $B_j$ into $(r-1)$ parts and index the subfiles using the elements of the set \mbox{$[r] \setminus \{j\}$}, i.e., the file $B_j$ is partitioned into subfiles $B_{j,1},B_{j,2},\dots,B_{j,j-1},B_{j,j+1},B_{j,j+2}\dots,B_{j,r}$.
We assume that each subfile is of size $\ell/(r-1)$.
In the protocol, each node $i$ broadcasts the following coded packet
\begin{equation*}
E_i = \bigoplus_{j \neq i} B_{j,i}
\end{equation*} 
to all the other nodes, where $\oplus$ denotes binary XOR. Since each coded packet is of length $\ell/(r-1)$ and there are $r$ such transmissions, the overall communication cost is $\ell\,r/(r-1)$.

We now argue that these $r$ coded packets are sufficient for each node to meet its demand. Consider a node $m$ that demands $B_m$ and observes the coded packets $E_i$, $i \neq m$. Note that the subfiles of $B_m$ are $B_{m,i}$, $i \neq m$. Node $m$ decodes the subfile $B_{m,i}$ from $E_i$ as follows
\begin{align*}
E_i \bigoplus \left( \bigoplus_{j \neq m,i} B_{j,i}  \right) &= \left( \bigoplus_{j \neq i} B_{j,i} \right) \bigoplus \left( \bigoplus_{j \neq m,i} B_{j,i}  \right) \\
&= B_{m,i}.
\end{align*}
This decoding operation is possible since node $m$ knows the files $B_j$, $j \neq m$. 

 \bibliographystyle{IEEEtran}
\bibliography{biblio.bib}
\end{document}